\documentclass{iaus}
\usepackage{graphicx}

\title[Spectra and Light Curves with SHELLSPEC.] 
{Synthetic Spectra and Light Curves of Interacting Binaries and Exoplanets
with Circumstellar Material: SHELLSPEC
}

\author[Jan Budaj]   
{Jan Budaj$^1$
}

\affiliation{$^1$Astronomical Institute, Tatransk\'{a} Lomnica, Slovakia, 
email: {\tt budaj@ta3.sk} \\[\affilskip]
}

\pubyear{2008}
\volume{xxx}  
\pagerange{119--126}
\jname{Title of your IAU Symposium}
\editors{A.C. Editor, B.D. Editor \& C.E. Editor, eds.}
\begin{document}

\maketitle

\begin{abstract}
Program SHELLSPEC is designed to calculate light-curves, spectra and images
of interacting binaries and extrasolar planets immersed in a moving
circumstellar environment which is optically thin. It solves simple
radiative transfer along the line of sight in moving media. The assumptions
include LTE and optional known state quantities and velocity fields in 3D. 
Optional (non)transparent objects such as a spot, disc, stream, jet, ufo,  
shell or stars may be defined (embedded) in 3D and their composite synthetic
spectrum calculated.
Roche model can be used as a boundary condition for the radiative transfer.
Recently a new model of the reflection effect, dust and Mie scattering
were incorporated into the code.

$\epsilon$ Aurigae is one of the most mysterious objects on the sky.
Prior modeling of its light-curve assumed dark, inclined,
disk of dust with the central hole to explain the light-curve with
a sharp mid-eclipse brightening. Our model consists of two geometrically 
thick flared disks. Internal optically
thick disk and external optically thin disk which absorbs and scatters 
radiation. Shallow mid-eclipse brightening may result from
eclipses by nearly edge-on flared (dusty or gaseous) disks.
Mid-eclipse brightening may also be due to strong forward scattering 
and optical properties of the dust which can have an important effect 
on the light-curves.

There are many similarities between interacting binary stars and transiting
extrasolar planets. Reflection effect which is briefly reviewed is one of
them. The exact Roche shape and temperature distributions over
the surface of all currently known transiting extrasolar planets have been
determined. In some cases (HAT-P-32b, WASP-12b, WASP-19b) departures from 
the spherical shape can reach 7-15\%.
\end{abstract}

\firstsection 

\section{Introduction}

There are sophisticated computer codes for calculating and inverting
light curves or spectra of binary stars with various shapes
or geometry including the Roche model 
(\cite[Lucy 1968]{lucy68}; \cite[Wilson \& Devinney 1971]{wd71}; 
\cite[Wood 1971]{wood71}; \cite[Mochnacki \& Doughty 1972]{md72}; 
\cite[Rucinski 1973]{rucinski73}; \cite[Hill 1979]{hill79}; 
\cite[Popper \& Etzel 1981]{pe81}; \cite[Djurasevic 1992]{djurasevic92}; 
\cite[Drechsel et al. 1994]{drechsel94}; \cite[Hadrava 1997]{hadrava97}; 
\cite[Bradstreet \& Steelman 2002]{bs02}; \cite[Pribulla 2004]{pribulla04}, 
\cite[Pavlovski et al. 2006]{pb06}, \cite[Tamuz et al. 2006]{tm06}).
The Wilson \& Devinney code is most often used and is continuously
being improved or modified (\cite[Kallrath et al.1998]{km98}; 
\cite[Pr\v{s}a \& Zwitter 2005]{pz05}).
The main focus of these codes is to deal with the stars,
determine their properties and their orbit. However, 
it is often the case that stellar objects are embedded in some moving
optically thin environment and/or are accompanied by discs, streams,
jets or shells which give rise to various emission spectra. 
We address these objects with Shellspec.

The Shellspec code is described in more detail in \cite[Budaj \& Richards
(2004)]{br04}. There has been a lot of progress
since that manual was published and an updated version of the manual,
with examples of input, output and test runs is available within each
new release. A convenient overview may also be found in 
\cite[Budaj \& Richards (2010)]{br10}.
Major changes since that time include a new model of the reflection effect
applicable to tidally distorted and strongly irradiated cold objects 
(\cite[Budaj 2011a]{budaj11a}), and dust and angle dependent Mie scattering 
(\cite[Budaj 2011b]{budaj11b}). It includes extinction due to 
the absorption and scattering as well as the thermal and scattering emission.
Original Shellspec code is written in Fortran77 and does not solve 
the inverse problem of finding the best fit parameters.
There are versions of this code in Fortran90 which even solve some
particular restricted inverse problems by 
\cite[Tkachenko et al. (2010)]{tkachenko10} and
\cite[\v{S}ejnov\'{a} et al. (2011)]{sejnova11}.
\cite{chadima2011a} modeled $H\alpha$ emission V/R variations 
caused by discontinuous mass transfer in interacting binaries.
Miller et al. (\cite{miller07}) modeled UV and optical spectra of TT Hya,
Algol type binary.
Sections below describe applications of this code to $\epsilon$ Aur 
and extrasolar planets.
Output of the codes SYNSPEC and COOL-TLUSTY
(\cite[Hubeny 1988]{hubeny88}; \cite[Hubeny, Lanz \& Jeffery 1994]{hlj94};  
\cite[Hubeny, Burrows \& Sudarsky 2003]{hbs03}) is used as a default input 
of spectra of nontransparent objects, which serves as a boundary condition 
for the radiative transfer in the interstellar medium.

\section{$\epsilon$ Aurigae}

$\epsilon$ Aur is an eclipsing binary with the longest known orbital 
period, 27.1 yr.
A very rare eclipse that lasted for two years is over.
However, the object, its origin and, in particular, the secondary
component of this binary star remain mysterious. 
Huang (\cite{huang}) proposed that the secondary is a dark disk seen
edge-on. Wilson (\cite{wilson}) and Carroll et al. (\cite{carroll}) 
argued that the observed sharp mid-eclipse brightening {\bf (MEB)} 
\footnote{By a mid-eclipse brightening in $\epsilon$ Aur some authors
understand only a relatively sharp local maximum that appeared in    
a few eclipses near the middle of the eclipse. We propose a slightly 
more general definition of the MEB or a new term (mid-eclipse excess).
Our MEB or mid-eclipse excess is a convex feature near the middle of  
an eclipse bed. A common eclipse has a concave eclipse bed.}
can only be explained by a tilted disk with a central opening.
Ferluga (\cite{ferluga}) suggested that the disk is a system of rings.

\begin{figure}[h]
\centerline{
\includegraphics[angle=-90,width=7.cm]{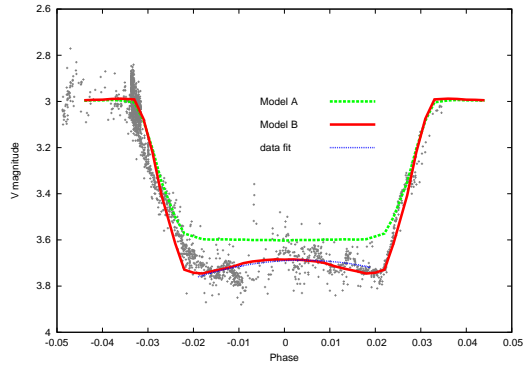}
}
\caption{
Eclipse of $\epsilon$ Aur by a dark, geometrically thick, flared
disk of dust. The disk consists of two parts:
(1) The flared optically thick part that causes most of the eclipse, and
(2) a flared optically thin part that causes additional absorption,
scattering and mid-eclipse brightening. 
Model (A): Disk has only one part (1). 
Model (B): disk has both parts part (1) and part (2).
Mid-eclipse brightening arises mainly because the edges of the flared disk
are more effective in the attenuation of the stellar light than the central
parts of the disk. Thin dotted line is a best-fit quadratic function to
the eclipse bed.
Crosses - observations from AAVSO (Henden \cite{henden}). 
}                                              
\label{epsaur1}   
\end{figure}

\begin{figure}[h]
\centerline{
\includegraphics[angle=0,width=8.cm]{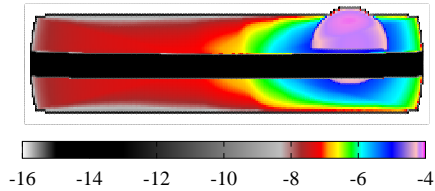}     
}
\vspace{-10mm}
\caption{
2D image of $\epsilon$ Aur during eclipse.
Black region is a dark, geometrically and
optically thick, edge-on flared disk of dust. This disk causes 
the most of the eclipse.
Colored regions correspond to the optically thin flared disk which
scatters and absorbs the light from the F-star.
Notice that this optically thin disk may produce artificial spots on the
surface of the F star.}                                              
\label{epsaur2}   
\end{figure}

There has been a wealth of studies during the current eclipse.
Orbital solutions were recently revisited by
Stefanik et al. (\cite{stefanik}) and Chadima et al. (\cite{chadima2010}).
Kloppenborg et al. (\cite{kloppenborg2010}, \cite{kloppenborg2011}) 
confirmed the dark disk with interferometric observations but 
they did not confirm the hole in the disk.
The spectral energy distribution was studied by Hoard et al. (\cite{hoard}).
These authors favor the post AGB+B5V model.
Wolk et al. (\cite{wolk}) analyzed X-ray observations.
Sadakane et al. (\cite{sadakane10}) carried out the abundance analysis.
Recently, Chadima et al. (\cite{chadima2010}, \cite{chadima2011b})
questioned the presence of sharp mid-eclipse brightening and 
suggested that the photometric variability seen during eclipse is
intrinsic to the F-star.
Extremely precise observation of the light-curve of $\epsilon$ Aur
were obtained with the Solar Mass Ejection Imager by 
\cite[Clover et al. (2011)]{clover11}.
These data clearly show a shallow mid eclipse brightening.

Observations that were used for comparison with our calculations
were taken from the AAVSO database (Henden \cite{henden}). They were
obtained by many observers who contributed to the database during 
the current eclipse of  $\epsilon$ Aur. Only the observations in 
the V-filter were considered here. 
These observations also indicate the presence of a shallow mid-eclipse      
brightening but it is not as sharp and pronounced as some might have         
anticipated.
Ingress is not as steep as egress, which indicates that 
the disk is not perfectly symmetric, but suffers from some small
disturbance. Its leading part might be more extended or disk slightly 
inclined out of the orbital plane (warped?).

To explain these observations we suggested an alternative model
of $\epsilon$ Aur.
Our model of $\epsilon$ Aur consists of two geometrically
thick flared disks: an internal optically thick disk and an external
optically thin disk, which absorbs and scatters radiation.
Disks are in the orbital plane and are almost edge-on.
The model is based on optical properties of dust grains.
It takes into account the extinction due to the Mie scattering and
absorption as well as thermal and scattering emission.
We argue that there is no need for a highly inclined
disk with a hole to explain the current eclipse of $\epsilon$ Aur
even if there is a possible shallow mid-eclipse brightening
(see Fig.\ref{epsaur1}).
For mode details kindly see \cite[Budaj (2011b)]{budaj11b}.
Fig.\ref{epsaur2} displays a 2D image of $\epsilon$ Aur
as calculated in the V band.

\section{Shapes of transiting extrasolar planets}

Transiting extrasolar planets are very close to their parent stars.
Most of them have circular orbits or very small eccentricity.
This indicates that their rotation is synchronous and classical
Roche model can be applied to them. \cite[Budaj (2011a)]{budaj11a}
calculated the shape of all transiting exoplanets known at that time.
By the shape we mean the relative proportions of the object.
Shape does not change much if the absolute dimensions change within
typical measurement errors. Consequently, the new absolute dimensions 
can be easily linearly rescaled e.g. if the measured cross-section of 
the planet changes. 
I use $R_{sub}/R_{pole}$ as a measure of the departure 
from the spherical shape, where $R_{sub}$ is the radius at the sub-stellar
point and $R_{pole}$ is the radius at the rotational pole.
I updated these calculations and recalculated the shape of all 138
transiting exoplanets known as of end of June, 2011.
Most of the transiting exoplanets have small departures from the sphere
of the order of 1\%. 
However, a fraction of them have departures more than 3\% and these
are listed in the Table 1.
HAT-P-32 b, WASP-12 b, and WASP-19 b are recorders with departures
of about 7, 12, and 15\% respectively.
This means that we are observing only a cross section of these planets 
during the transit. This cross-section is also not spherical, it is
characterized by the $R_{side}/R_{pole}$ parameter but this is not as high
as the $R_{sub}/R_{pole}$. Table 1 lists also the effective radius of 
the planet (radius of the sphere with the same volume) which one could use 
e.g. for comparison with the theoretical radius calculations.
Table 1 illustrates that spherical shape of close-in exoplanets cannot be 
taken for granted. Moreover, the Table 1 demonstrates that this Roche shape 
cannot be approximated by the {\it rotational} ellipsoid.
This assumption used in some papers is not justified for most of the
transiting extrasolar planets.

\begin{table}
\begin{center}
\caption{
Shapes of the transiting exoplanets.
Columns are:
$a$ -semi-major axis in [AU], 
$R_{sub}$  -planet radius at the sub-stellar point,
$R_{back}$ -planet radius at the anti-stellar point,
$R_{pole}$ -planet radius at the rotation pole,
$R_{side}$ -planet radius at the side point,
(assumed equal to the planet radius determined from the transit),
$R_{eff}$  -effective radius of the planet,
$rr=R_{sub}/R_{pole}$ -departure from the sphere,
$f_{i}=R_{sub}/L_{1x}$ -fill-in parameter of the Roche lobe,
Radii are in units of Jupiter radius. 
See the text for a more detailed information.
\label{T1}
}
\begin{tabular}{lllllllll}
\hline
 a      &  Rsub   & Rback   & Rpole   & Rside   &Reff    & rr    & $f_{i}$ & name \\
\hline 
 0.01655 & 1.55474 & 1.54069 & 1.35022 & 1.38600 & 1.42045 & 1.151 &0.63 &  WASP-19 b    \\
 0.02293 & 1.90438 & 1.89283 & 1.69728 & 1.73600 & 1.77110 & 1.122 &0.58 &  WASP-12 b    \\
 0.03440 & 2.14496 & 2.14005 & 2.00804 & 2.03700 & 2.06019 & 1.068 &0.48 &  HAT-P-32 b   \\
 0.02540 & 1.54395 & 1.54167 & 1.47462 & 1.49000 & 1.50166 & 1.047 &0.42 &  CoRoT-1 b    \\
 0.02312 & 1.40758 & 1.40572 & 1.35013 & 1.36300 & 1.37264 & 1.042 &0.41 &  WASP-4 b     \\
 0.05150 & 2.04439 & 2.04295 & 1.97515 & 1.99100 & 2.00266 & 1.035 &0.38 &  WASP-17 b    \\
 0.02250 & 1.23086 & 1.22972 & 1.19084 & 1.20000 & 1.20670 & 1.033 &0.37 &  OGLE-TR-56 b \\
 0.03444 & 1.70957 & 1.70825 & 1.65814 & 1.67000 & 1.67862 & 1.031 &0.36 &  WASP-48 b    \\
\hline
\end{tabular}
\end{center}
\end{table}

\section{Reflection effect}

Reflection effect operates in very different environments and there are
different approaches to model this effect.
The table below presents an overview of how the models work in the field of
interacting binaries and exoplanets.
One can see that the models of the reflection effect and definition of
the fundamental quantities are very different in these fields.
The question is: Is there any transition between the two approaches?
What is the amount of the heat redistribution? What fraction of light
gets reflected off the surface and does not produce heat? 
Which approach (model) 
should one choose for tidaly distorted and strongly irradiated cold object?
To address these questions and these objects \cite[Budaj (2011a)]{budaj11a}
proposed a sort of hybrid model of the reflection effect whose properties 
are also listed in the table.

\begin{table}
\begin{center}
\caption{
Overview of the reflection effect.
Scattering refers to the refletion of light off the surface of one of the
objects. $A_{bol}$ is bolometric albedo (\cite[Rucinski 1969]{rucinski69}), 
$A_{B}$ is Bond albedo, $A_{\nu}$ is monochromatic albedo.
MR stands for muliple reflection between surfaces of the two objects.
\label{T2}
}
\begin{tabular}{llll}
\hline
 Property     & Interacting Binaries    &  Exoplanets            &   Hybrid model           \\
\hline 
Shape         & Roche (limb+grav.dark.) & sphere, rot. ellipsoid & Roche (limb+grav.dark.)  \\
Scattering    & No                      & Yes                    & Yes                      \\
Albedo        & $A_{bol}$               & $A_{B}, A_{\nu}$       & $A_{B}, A_{\nu}$         \\
(A) Meaning   & absorbed$=>$heating     & reflected$=>$no heating & reflected$=>$no heating \\
(1-A) Meaning & penetrates into star    & heating+heat redistrib. & heating+heat redistrib. \\
MR            & Yes                     & No                      & No                      \\ 
\hline
\end{tabular}
\end{center} 
\end{table}



\acknowledgements
JB thanks the VEGA grants No.s 2/0074/9, 2/007810, 2/0094/11.

\begin{discussion}

\discuss{Wilson}{
Did you look into the possible importance of multiple scattering?  The 
scattering geometry affects the angular distribution of scattered light 
and, to a small extent, even whether a scattered photon escapes the 
atmosphere.  The effect is small but might not be negligible.
}

\discuss{Budaj}{
There is a scattering (reflection) process in two environments.
One is the scattering in the atmosphere of stars or nontransparent objects.
I do not solve the radiative transfer in the atmospheres of these objects.
I take the flux from the model atmosphere calculation and use is to
calculate the boundary condition for the radiative transfer in the moving 
interstellar matter. Models of the atmosphere take the multiple scattering 
into account.
The other is the scattering process in the interstellar medium.
In this case I assume that the medium is optically thin and that
it is irradiated by one or two sources. I take into account only the first
scattering event. In the optically thin medium the probability of the second
and higher scattering events rapidly decreases. To take them into account
one would have to solve the 3D radiative transfer in the 3D moving medium.
}

\discuss{Burrows}{
Have you thought of using your code to calculate light curves for 
WASP-12b at the warm Spitzer bands (3.6 μm and 4.5 μm), instead of at 8 
μm?  There should be some data soon from Spitzer for WASP-12b at those 
bands and such models would be relevant.
}

\discuss{Budaj}{
No. However, this would be indeed very interesting. Thank you for the
suggestion/information.
}

\discuss{Rucinski}{
I congratulate you for making order with definitions of albedo, as 
applied to stars and planets.  The old definition was directly related 
to the so-called “reflection effect” in eclipsing binaries.  
}

\discuss{Budaj}{
Thank you. 
}

\discuss{Southworth}{
You find that the difference in radius between a spherical planet and 
the substellar point of a Roche-model planet is as large as 15\%.  When a 
planet is transiting we do not see its substellar point but we do 
measure its equatorial/ polar radius.  This will be closer to the radius 
of a spherical planet so the problem is not as bad.  Can you put a 
figure on this?
}

\discuss{Budaj}{
Yes. That is true (if one deals with transits and not with other
phases). $R_{side}/R_{pole}$ is not as big. I do not remember the exact 
numbers but they are in my paper for all the transiting planets. 
This ratio might be about 3\% in the worst case. The effective radius of 
the planet (radius of the sphere with the same volume) is also tabulated 
there.
}

\end{discussion}


\begin{thebibliography}{}

\bibitem[2002]{bs02}
Bradstreet, D.H., \& Steelman, D.P. 2002, Bull. Am. Astron. Soc., 34, 1224    

\bibitem[Budaj (2011a)]{budaj11a}
Budaj, J. 2011a, AJ, 141, 59

\bibitem[Budaj (2011b)]{budaj11b}                
Budaj, J. 2011b, A\&A, 532, L12

\bibitem[2004]{budaj2004}
Budaj, J., \& Richards, M.T. 2004,
Contrib. Astron. Obs. Skalnat\'{e} Pleso, 34, 167

\bibitem[Budaj \& Richards (2010)]{br10}
Budaj, J., \& Richards, M.T. 2010, 
ASP Conf. Ser. 435, 63

\bibitem[1991]{carroll}
Carroll, S.M., Guinan, E.F., McCook, G.P., \& Donahue, R.A.
1991, ApJ, 367, 278

\bibitem[2010]{chadima2010}
Chadima, P., Harmanec, P., Yang, S. et al.
2010, IBVS, 5937

\bibitem[Chadima et al. (2011a)]{chadima2011a}
Chadima, P., Firt, R., Harmanec, P. et al.
2011a, AJ, 142, 7

\bibitem[2011b]{chadima2011b}
Chadima, P., Harmanec, P., Bennett, P.D. et al.
2011b, A\&A, 530, A146

\bibitem[2011]{clover2011}	
Clover, J., Jackson, B. V., Buffington, A., Hick, P. P.,
Kloppenborg, B., Stencel, R. 2011,
AAS Meeting 217, 257.02

\bibitem[1992]{djurashevic92}
Djurasevic, G., 1992, Astrophysics and Space Science, 197, 17

\bibitem[1994]{dh94}
Drechsel, H., Haas, S., Lorenz, R., \& Mayer, P. 1994, A\&A, 284, 853

\bibitem[1990]{ferluga}
Ferluga, S. 1990, A\&A, 238, 270

\bibitem[1997]{hadrava97}
Hadrava, P. 1997, A\&AS, 122, 581

\bibitem[2011]{henden}
Henden, A.A., 2011, Observations from the AAVSO International Database,
private communication.

\bibitem[1979]{hill79}
Hill, G. 1979, Publ. Dom. Ap. Obs. Victoria, 15, 297

\bibitem[2010]{hoard} 
Hoard, D. W., Howell, S. B., \& Stencel, R. E.
2010, ApJ, 714, 549

\bibitem[Hubeny (1988)]{hubeny88}
Hubeny, I. 1988, 
Computer Physics Comm., 52, 103

\bibitem[Hubeny, Burrows \& Sudarsky (2003)]{hbs03}
Hubeny, I., Burrows, A., \& Sudarsky, D. 2003, 
ApJ, 594, 1011

\bibitem[Hubeny, Lanz \& Jeffery (1994)]{hlj94}
Hubeny, I., Lanz, T., Jeffery, C.S.: 1994, in Newsletter on Analysis
of Astronomical spectra No.20, ed. C.S. Jeffery (CCP7; St. Andrews:
St. Andrews Univ.), 30

\bibitem[1965]{huang}
Huang, S. 1965, ApJ, 141, 976

\bibitem[1998]{km98}
Kallrath, J., Milone, E.F., Terrell, D., \& Young, A.T.
1998, ApJ, 508, 308

\bibitem[2010]{kloppenborg2010}
Kloppenborg, B., Stencel, R., Monnier, J.D. et al.
2010, Nature, 464, 870

\bibitem[2011]{kloppenborg2011}
Kloppenborg, B.K., Stencel, R., Monnier, J.D. et al.
2011, AAS Meeting 217, 257.03


\bibitem[1968]{lucy68}
Lucy, L.B. 1968, ApJ, 153, 877

\bibitem[2007]{miller07}
Miller, B., Budaj, J., Richards, M., Koubsk\'{y}, P., \& Peters, G. 
2007, ApJ, 656, 1075

\bibitem[1972]{md72}
Mochnacki, S.W., \& Doughty, N.A., 1972, MNRAS, 156, 51

\bibitem[2006]{pb06}
Pavlovski, K., Burki, G., \& Mimica, P. 2006, A\&A, 454, 855

\bibitem[1981]{pe81}
Popper D.M., \& Etzel P.B. 1981, AJ, 86, 102

\bibitem[2004]{pribulla04}
Pribulla, T. 2004, Spectroscopically and Spatially
Resolving the Components of Close Binary Stars,
R. W. Hidlitch, H. Hensberge \and K. Pavlovski,
ASP Conf. series., 318, 117

\bibitem[2005]{pz05}
Pr\v{s}a, A., \& Zwitter, T. 2005, ApJ, 628, 426

\bibitem[Rucinski(1969)]{rucinski69}
Rucinski, S.M. 1969, Acta Astronomica, 19, 245

\bibitem[1973]{rucinski73}
Rucinski, S.M. 1973, Acta Astronomica, 23, 79

\bibitem[2010]{sadakane10}
Sadakane, K., Kambe, E., Sato, B., Honda, S., Hashimoto, O.
2010, PASJ, 62, 1381

\bibitem[2010]{stefanik}
Stefanik, R.P., Torres, G., Lovegrove, J. et al.
2010, AJ, 139, 1254


\bibitem[\v{S}ejnov\'{a} et al. (2011)]{sejnova11}
\v{S}ejnov\'{a}, K., Votruba, V., \& Koubsk\'{y}, P. 2011, this proceedings

\bibitem[2006]{tm06}
Tamuz O., Mazeh T., \& North P. 2006, MNRAS, 367, 1521

\bibitem[Tkachenko et al. (2010)]{tkachenko10}
Tkachenko, A., Lehmann, H., \& Mkrtichian, D. 2010, AJ, 139, 1327

\bibitem[1971]{wilson}
Wilson, R.E. 1971, ApJ, 170, 529

\bibitem[1971]{wd71}
Wilson, R.E., \& Devinney, E.J. 1971, ApJ, 166, 605

\bibitem[2010]{wolk}
Wolk, S.J., Pillitteri, I., Guinan, E., \& Stencel, R.
2010, AJ, 140, 595

\bibitem[1971]{wood71}
Wood, D.B. 1971, AJ, 76, 701






\end{thebibliography}
\end{document}